\documentstyle[11pt,paspconf,epsf,psfig,twoside]{article}

\newcommand{\rx}{\mbox{RX\,J0803.4--4748}}
\newcommand{\rxs}{\mbox{0803}}
\def\degr{\hbox{$^\circ$}}
\def\arcmin{\hbox{$^\prime$}}
\def\arcsec{\hbox{$^{\prime\prime}$}}
\def\fd{\hbox{$.\!\!^{\rm d}$}}
\def\fm{\hbox{$.\!\!^{\rm m}$}}

\begin{document}


\title{Filling up the period gap: RX\,J0803.4--4748, a new 137 minute polar} 

\author{Robert Schwarz  and Jochen Greiner}

\affil{Astrophysical Institute Potsdam, 14482 Potsdam, Germany}
\begin{abstract}
We report the discovery of a new AM Herculis system as the counterpart
of the bright and very soft X--ray source \rx\
found serendipitously in a ROSAT PSPC pointing. 
Optical photometry showed deep (2 mag)
variations with a period of $\sim 137$~min placing \rx\ right into the period 
gap. The modulation can be interpreted in terms of cyclotron beaming and 
is very similar to that of other non--selfeclipsing polars like e.g. V843~Cen.
Such a geometry is also supported by the X--ray light curve of a
later HRI pointing which was lacking any pronounced faint--phase.
Phase-resolved optical spectra are charaterized by strong Balmer and 
He\,{\sc ii} $\lambda 4686$\AA\ line emission
superimposed onto a blue continiuum, and reveal
disinct cyclotron humps at certain phases. The 
field strength of 39 MG as derived from the
separation of the harmonics is in agreement with that of other soft
X--ray
polars.
\end{abstract}


\section{ROSAT observations}

\begin{figure}
\mbox{
\psfig{figure=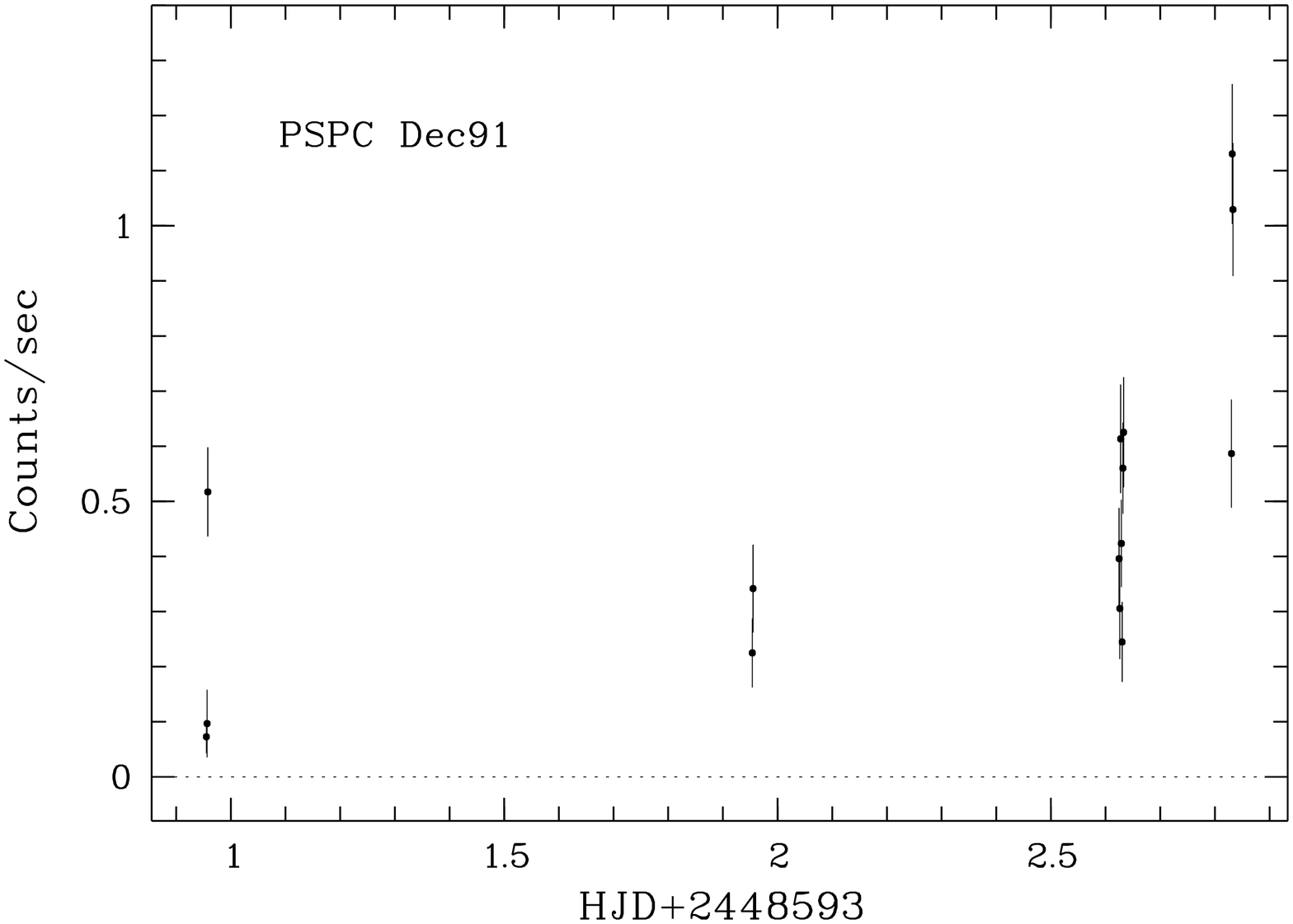,angle=-90,width=6.3cm,clip=}
\psfig{figure=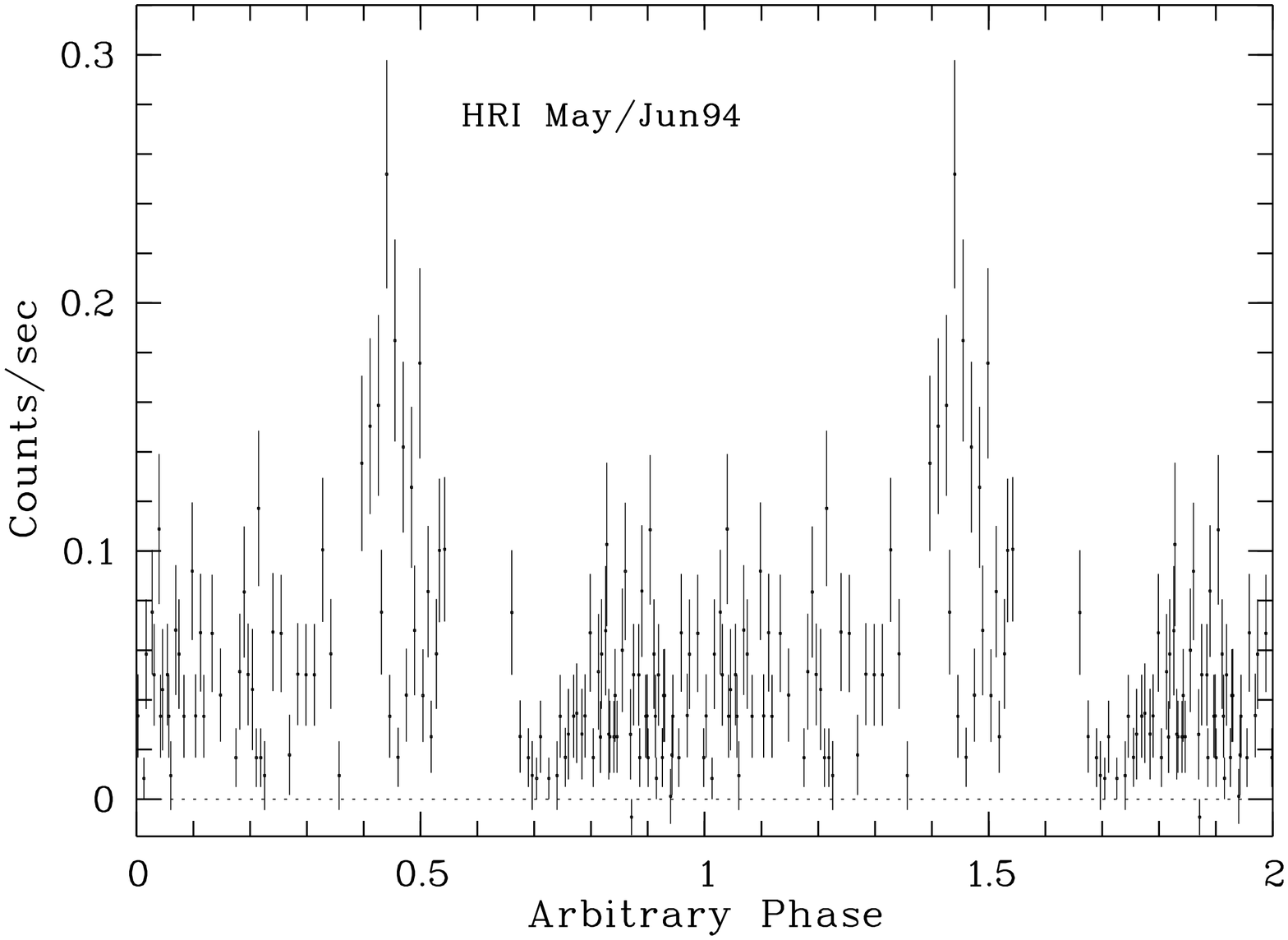,angle=-90,width=6.3cm,clip=}
}
\caption{ {\bf Left:}
X-ray light curve of \rx\ obtained on December 3, 1991 
   using the ROSAT PSPC. {\bf Right:}
The HRI X-ray light curve obtained in May/June 1994 
folded with a period of 137.1 min. The phase zero offset is arbitrary.} 
\label{fig1}
\end{figure}

As part of a program to study newly ROSAT-discovered supersoft 
X-ray sources (Greiner 1996) the field of \rx\ was observed between 
December 2--4, 1991 in a short (2.3 ksec) pointing centered on 
RX\,J0800.4--4746 which later turned out to be a white dwarf (Fleming et al.
1996).  \rx\ ($\equiv$1RXS 080346.3--474838; henceforth referred to as \rxs) 
appeared therein as the second brightest X-ray source 
(average count rate $0.465 \pm 0.016$~cts/sec) located $\sim$37\arcmin\  
off-axis. It's X-ray flux was strongly modulated ($\sim$95\%)
rising to a peak-rate of  $\sim$1.2 cts/sec at the end of the pointing 
(Fig.~\ref{fig1}; left).
The X-ray spectrum is very soft, with virtually no counts above 0.5~keV as 
indicated by the hardness ratio HR1=$-0.93\pm0.07$. The spectrum can be readily
deconvolved with an absorbed blackbody model ($\chi^2_{\nu}$=1.04). 
The inclusion of an additional bremsstrahlung component (with fixed $T_{\rm
br}=20$ keV) yields no significantly better fit ($\chi^2_{\nu}$=1) 
and only slightly alters the blackbody parameters. The best-fit range of 
temperature and absorbing columns is
$T_{\rm bb}=30-55$~eV and $N_{\rm H}=(2-8)\cdot 10^{20}$~cm$^{-2}$
at the 95\% confidence level. 
We can set a firm upper limit on the flux-ratio in the ROSAT band 
$F_{\rm br}/F_{\rm bb}\la 0.01$
(with $T_{\rm bb}\equiv 25$~eV and $T_{\rm bb}\equiv  20$~keV fixed).

The positional error of $\sim$50\arcsec\ made an optical identification in 
this crowded galactic plane region (b$^{\mbox{\sc ii}}$=--8\fdg7) difficult. 
We therefore obtained a ROSAT HRI pointing between May 12 and June 7, 1994
which constrained the X-ray position to 
$\alpha_{2000} = 08^{\rm h}03^{\rm m}45\fs 6$ 
and $\delta_{2000}=-47\degr 48\arcmin 42\arcsec$. The CV is the 
only optical counterpart found within the 10\arcsec\ HRI error circle 
and has end figures 45\fs6 and 47\arcsec.
The X-ray light-curve of the HRI observation 
folded on the preliminary photometric ephemeris (see
below) revealed no strong orbital variations, but a non-zero count-rate at
nearly all phases with an average count rate of $0.045 \pm 0.007$
(Fig.~\ref{fig1}; right)
consistent with the mean rate during the PSPC pointing (note the conversion
factor of 7.8 between HRI and PSPC count rates for such soft X-ray spectra).  
only one observing interval and is thus certainly not a permanent feature.

\section{Optical Photometry}
\begin{figure}[t]
\begin{minipage}[b]{6.3cm}
      \psfig{figure=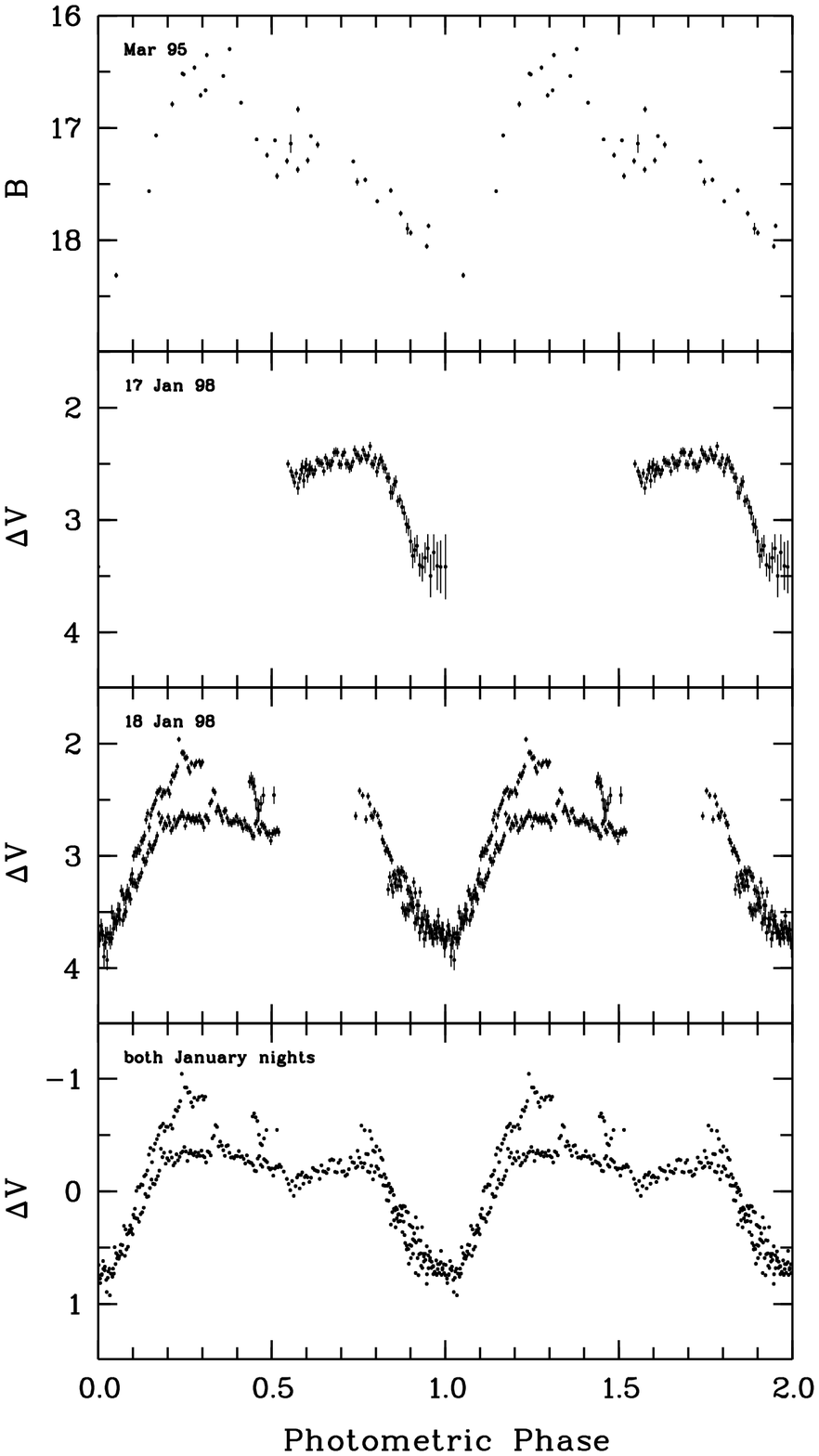,width=6.2cm,bbllx=50pt,bblly=90pt,%
       bburx=450pt,bbury=770pt,clip=}
\end{minipage}
\begin{minipage}[b]{6.3cm}
      \psfig{figure=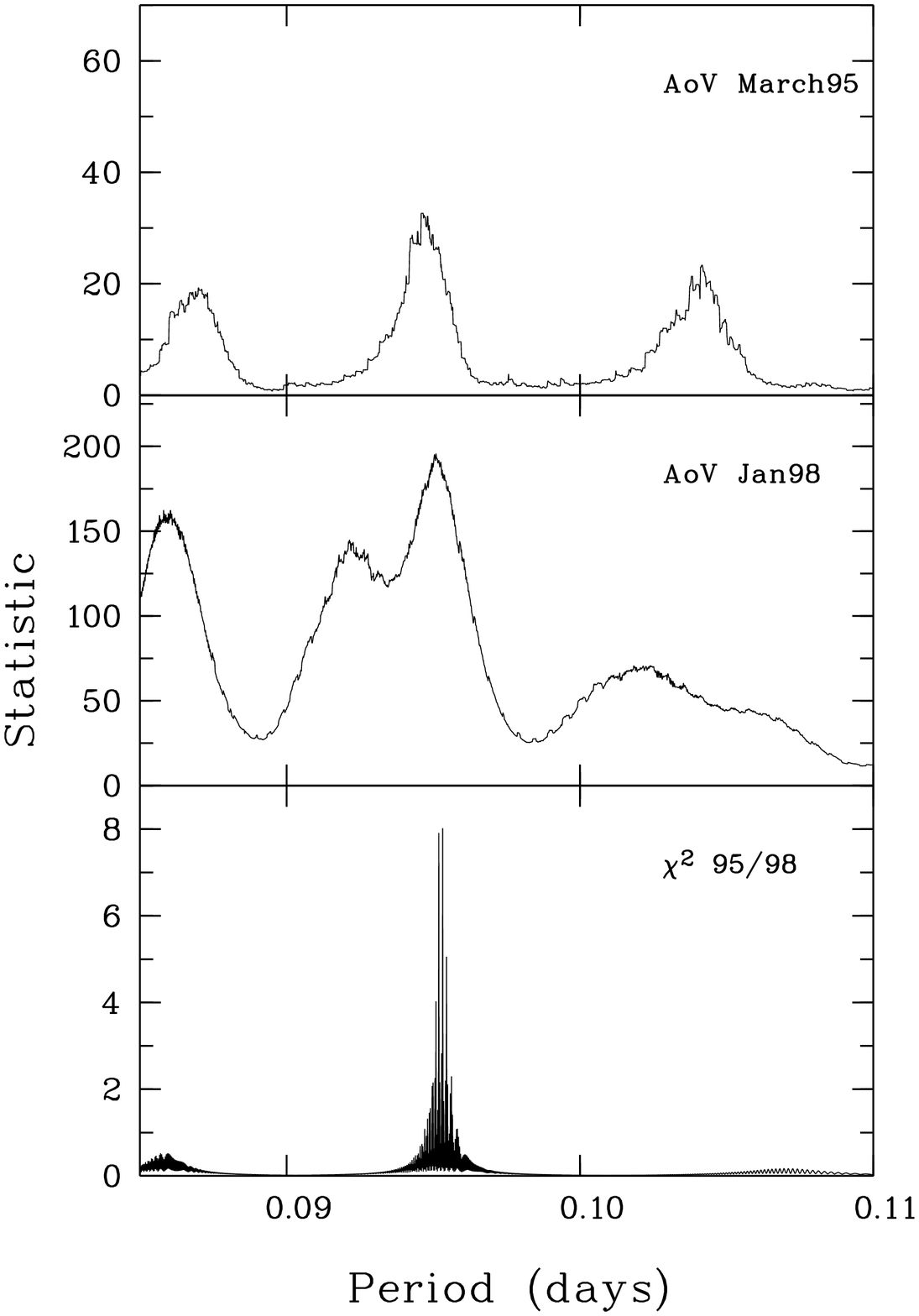,width=6.2cm,bbllx=30pt,bblly=45pt,%
       bburx=545pt,bbury=790pt,clip=}
\end{minipage}
\caption{{\bf Left:} Light curves of \rx\ obtained in March 1995 and 
  January 1998 folded with a period of 137.1 min. {\bf Right} 
  AoV-periodograms calculated from the optical photometry using the data from 
  March 1995 (upper panel) and January 1998 (middle panel). The lower panel 
  shows a weighted $\chi^2_{\nu}$-calculation applied to the timings of 
  the minima of both runs.} 
\label{fig2}
\end{figure}
Follow-up CCD-photometry was obtained on March  26-27, 1995 and January
17-18, 1998 at La Silla using the  2.2~m+EFOSC and the
1.5~m-Danish+DFOSC, respectively. We carried out a
period search on the individual data sets
using the analysis-of-variance method (Schwarzenberg-Cerny 1989).
Both  periodograms (Fig.~\ref{fig2}; right) are consistent with
a best-fit period of $\sim$137 min, although one-day aliases at 122
and 145 min are quite prominent. The latter are definitively ruled
out by a $\chi^{2}$ calculation applied to the timings of the four minima 
observed (Fig.~\ref{fig2}; right, lower panel). 
This shows a fine alias structure (due to the 3 year separation) 
around the best guess period at 137.1 min 
which is used to define a preliminary ephemeris of
\begin{equation}
{\rm HJD} ({T_{\rm min}}) = 245\, 0832.7476 + E \times 0.09523.
\end{equation}
Due to the aliasing the estimated uncertainty is still high ($\pm$0\fd 0003). 

The folded light curves (Fig.~\ref{fig2}; left) are characterized by a V-shaped minimum 
with 2 mag amplitude ($B = 16\fm2-18\fm5$) followed by a double-humped  
bright phase.  The overall appearance is similar to that of V834 Cen 
(Schwope \& Beuermann 1990), suggesting a geometry where the main accreting 
pole is permanently in view ($i+\beta < 90$\degr) and closest to the observer
at optical minimum. Accretion rate and/or geometry may vigorously change
on rather small time scales causing the 
large cycle-to-cycle variations (of about 0.5 mag) 
as observed on January 18, 1998.

\section{Optical Spectroscopy}
\begin{figure}[t]
\begin{minipage}[b]{6.3cm}
    \psfig{figure=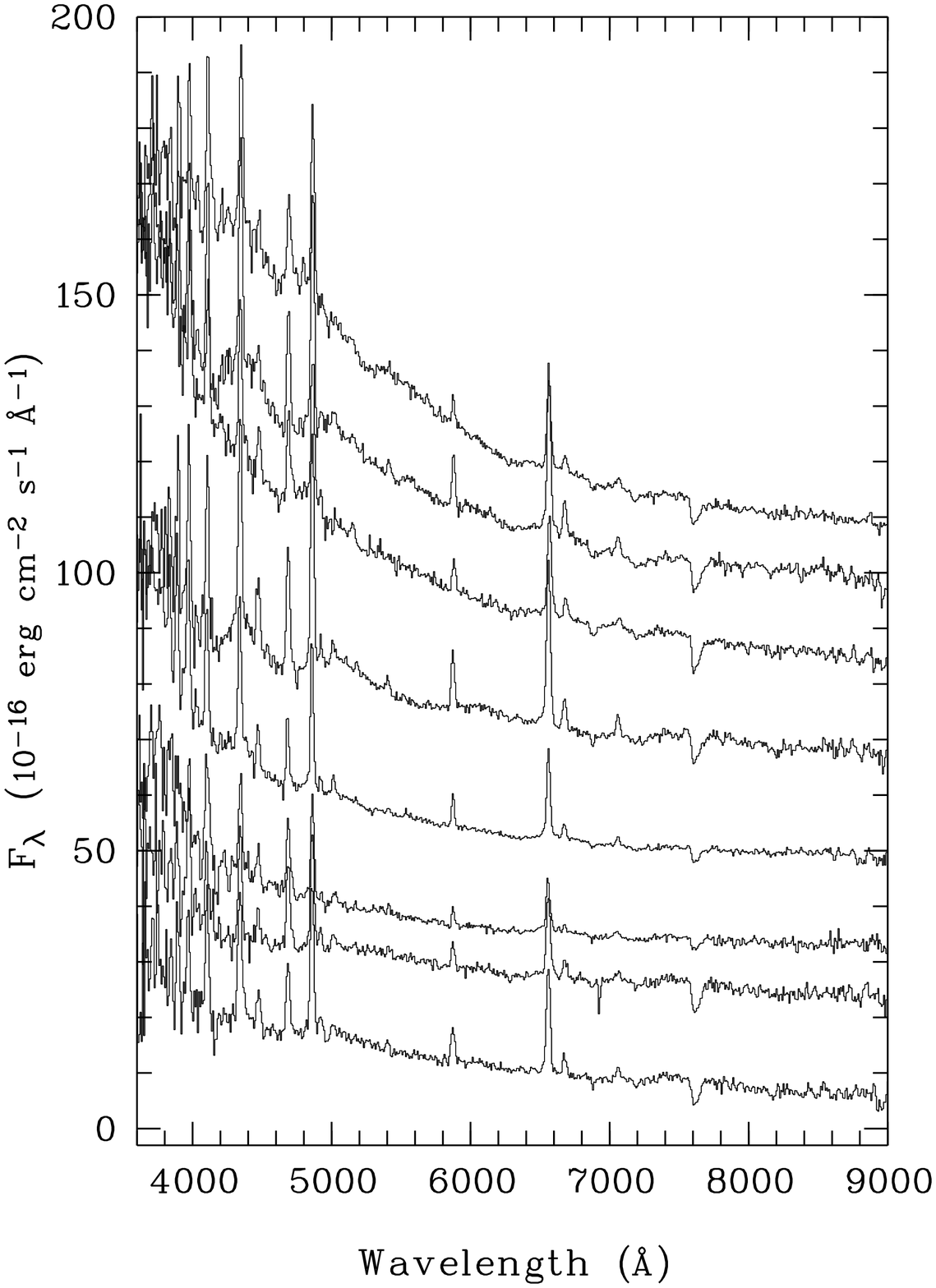,width=6.3cm,clip=}
\end{minipage}
\begin{minipage}[b]{6.3cm}
    \psfig{figure=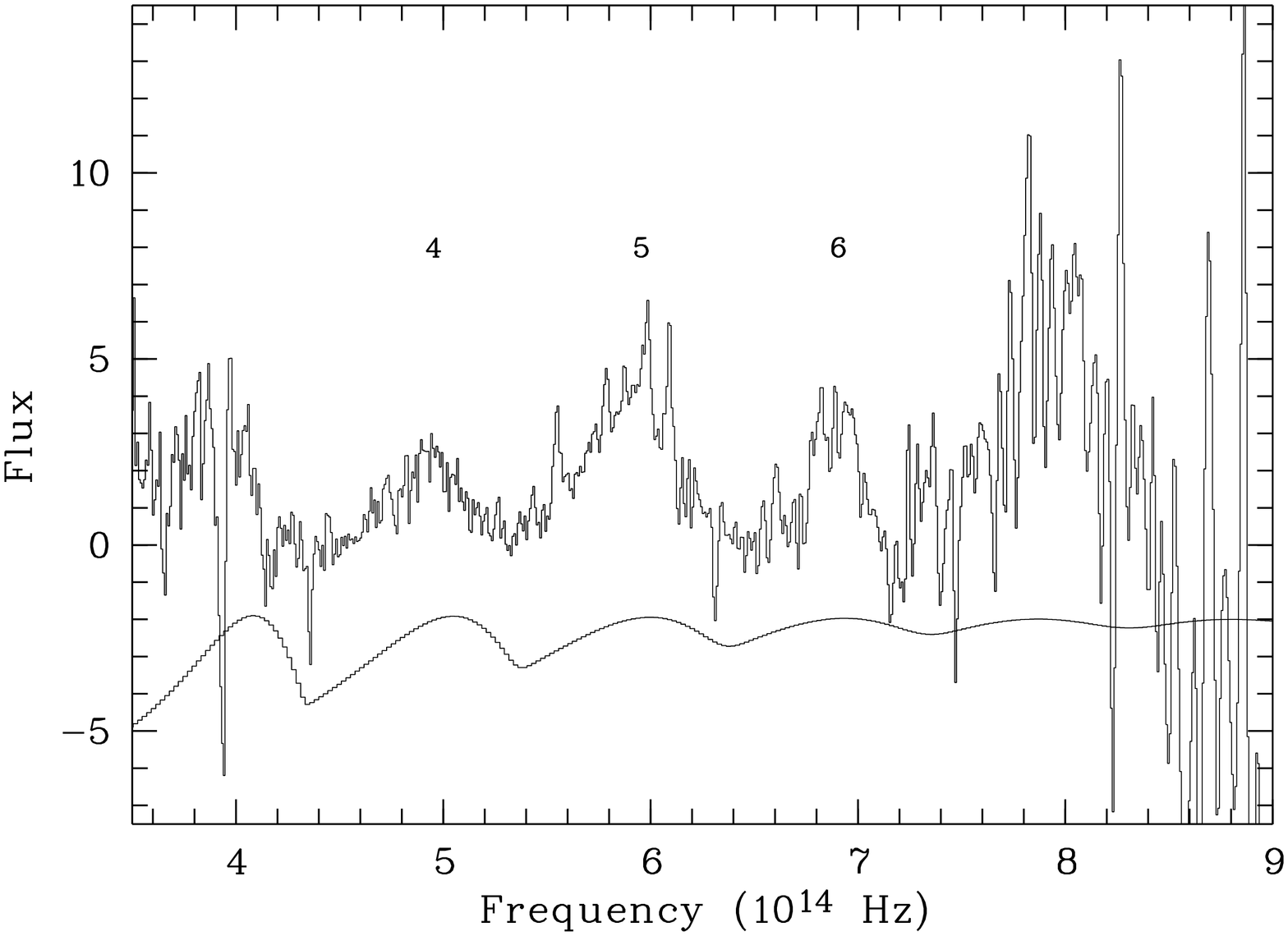,angle=-90,width=6.3cm,clip=}
    \rule{0pt}{16pt}
\end{minipage}
\caption{{\bf Left:} 
  Phase resolved spectra of \rxs\ obtained in March 1995.
  Spectra are shifted by 10 flux units each and run from phase 
  $\sim$0.0 (bottom) to $\sim$0.95 (top). 
 {\bf Right:} 
    Continuum and emission line subtracted spectrum taken at phase
    0.82 together with a cyclotron model of $B=39\pm3$ MG with $kT=10$ keV 
    and $\theta=80$\degr.
   }
\label{fig3}
\end{figure}

Our 1995 photometry was complemented by 8 low-resolution spectra 
covering approximately one orbital cycle with a time resolution of 15 min. 
Fig.~\ref{fig3} (left)  shows a subset of consecutive spectra.
The spectra are dominated by intense
emission lines of the Balmer series, He\,{\sc ii} $\lambda 4686$\AA,
and He\,{\sc i} superimposed on a blue continuum. The inverted Balmer
decrement and the strength of the  He\,{\sc ii} $\lambda 4686$\AA\
line being about 2/3 of H$\beta$ suggest a magnetic CV classification. 
This is directly confirmed by the detection of cyclotron lines at phases
0.7--0.8. The separation of the 4th/5th/6th harmonics (seen at 
$\lambda 6100$, $\lambda 5000$ and $\lambda 4300$\AA) of the cyclotron 
fundamental suggest a magnetic field strength of 39$\pm$3 MG
depending on the plasma temperature and the viewing angle $\theta$.
A corresponding cyclotron model 
is shown in Fig.~\ref{fig3} (right) together with the continuum and emission line subtracted
spectrum taken at phase 0.82.

\acknowledgments
We thank R. Egger for help during the 1995 ESO observing run.
JG and RS are supported by the Deut\-sches Zentrum f\"ur
Luft- und Raumfahrt (DLR) GmbH under contract No. FKZ 50 QQ 9602\,3 and
50 OR 9206\,8. The ROSAT project
is supported by the German Bundes\-mini\-ste\-rium f\"ur Bildung, Wissenschaft,
For\-schung und Technologie (BMBF/DLR) and the Max-Planck-Society.

%
%

\end{document}